\PassOptionsToPackage{unicode}{hyperref}
\PassOptionsToPackage{hyphens}{url}
\documentclass[11pt]{article}
\makeatletter
\g@addto@macro\@openbib@code{\setlength{\itemsep}{-5pt}}
\makeatother
\usepackage{amsmath,amssymb}
\usepackage{lmodern}
\usepackage{iftex}
\ifPDFTeX
  \usepackage[T1]{fontenc}
  \usepackage[utf8]{inputenc}
  \usepackage{textcomp} 
\else 
  \usepackage{unicode-math}
  \defaultfontfeatures{Scale=MatchLowercase}
  \defaultfontfeatures[\rmfamily]{Ligatures=TeX,Scale=1}
\fi
\IfFileExists{upquote.sty}{\usepackage{upquote}}{}
\IfFileExists{microtype.sty}{
  \usepackage[]{microtype}
  \UseMicrotypeSet[protrusion]{basicmath} 
}{}
\makeatletter
\@ifundefined{KOMAClassName}{
  \IfFileExists{parskip.sty}{%
    \usepackage{parskip}
  }{
    \setlength{\parindent}{0pt}
    \setlength{\parskip}{6pt plus 2pt minus 1pt}}
}{
  \KOMAoptions{parskip=half}}
\makeatother
\usepackage{xcolor}
\newlength{\cslhangindent}
\newlength{\csllabelwidth}
\newlength{\cslentryspacingunit} 
 {
  \setlength{\parindent}{0pt}
  \ifodd #1
  \let\oldpar\par
  \def\par{\hangindent=\cslhangindent\oldpar}
  \fi
  \setlength{\parskip}{#2\cslentryspacingunit}
 }%
 {}
\usepackage{calc}

\ifLuaTeX
  \usepackage{selnolig}  
\fi
\IfFileExists{bookmark.sty}{\usepackage{bookmark}}{\usepackage{hyperref}}
\IfFileExists{xurl.sty}{\usepackage{xurl}}{} 
\usepackage{fullpage}
\ifPDFTeX
\DeclareUnicodeCharacter{03BB}{$\lambda$}
\DeclareUnicodeCharacter{221E}{$\infty$}
\fi
\hypersetup{
  pdftitle={Wave-function parametrization of a probability measure},
  pdfauthor={Leonardo Pedro},
  hidelinks,
  pdfcreator={LaTeX via pandoc}}

\title{Wave-function parametrization of a probability measure}
\author{Leonardo Pedro}
\date{\today}

\begin{document}
\maketitle
\begin{abstract}
We show that any conditional probability measure in a standard measure
space is parametrized by a unitary operator on a separable Hilbert
space. We propose unitary inference, a generalization of Bayesian
inference. We study implications for classical statistical mechanics and
machine learning.
\end{abstract}

\hypertarget{introduction}{%
\section{1. Introduction}\label{introduction}}

In Scientific Research as in Bayesian inference, there is always a prior
probability distribution, and there is no prior which is better for all
cases\cite{dutchbook}: we
always have to make assumptions. For instance, if we insist we can
assume all objects in the night sky revolve around the Earth using
epicycles and even consider that this is a very successful theory
because it was a precursor of Fourier series and these have widespread
applications today\cite{acosta2019epicycles}.

\begin{quote}
\emph{The Emperor realized that the people were right but could not
admit to that. He though it better to continue the procession under the
illusion that anyone who couldn't see his clothes was either stupid or
incompetent. And he stood stiffly on his carriage, while behind him a
page held his imaginary mantle.}

---Hans Christian Andersen (1837)
\end{quote}

Thus, data, evidence and even mathematical proofs are not enough for us
to abandon prior beliefs, there must be a moment when in order for us to
achieve something we want we have to abandon our prior beliefs: in the
case of assuming all objects in the night sky revolve around the Earth,
it would make many relevant calculations in modern astronomy or in the
Global Positioning System impossible.

In the case of Quantum Mechanics, it was always possible to see its
mathematical formalism as a mere (but very useful) parametrization of
probability and classical information
theories\cite{Spekkens_2007}. But, it
was also possible to describe it (although in a forceful way) as a
\emph{new} and exotic alternative to (or generalization of) classical
information theory, one that shook our sense of reality. It is not hard
to believe that the mystery surrounding the exotic point of view can
often convince the society to spend large amounts of money in research,
at least in the short term.

The exotic point of view was sustained by a relatively constant flow of
discoveries of new phenomena at higher and higher energies, which kept
the mystery alive. But now, even optimists say that if the Large Hadron
Collider finds nothing new, it will be harder to convince the
governments of the world to build the next bigger, more expensive
collider to keep research in High Energy Physics as we know
it\cite{nightmare}. Thus, if the exotic
point of view is not that mysterious nowadays and there exists a
parametrization point of view, is there another way to convince the
society to spend large amounts of money in research?

The answer is yes, there is another way. Because the parametrization
point of view allows the mathematical formalism of Quantum Mechanics to
be applied to probability and classical information theory (and then to
artificial intelligence) not just to edge cases, but to the core of
these theories. This would not be possible with the exotic point of
view, which considered that Quantum Mechanics shook our sense of reality
and thus must be an alternative to (or a generalization of) classical
information theory and thus there would be no reason to expect
applications to the core of classical information theory.

\hypertarget{probability-updates-machine-learning-and-quantum-mechanics}{%
\section{2. Probability updates, machine learning and Quantum
Mechanics}\label{probability-updates-machine-learning-and-quantum-mechanics}}

We call to the process of updating a state of knowledge with respect to
new evidence as Probability
updates\cite{VanFraassen1993}.
There is little doubt that the Bayes rule should be applied to
Probability updates, \emph{whenever
possible}\cite{VanFraassen1993}.
But there is also little doubt that the Bayes rule is often not
applicable\cite{VanFraassen1993}
and thus Probability updates is a generalization of Bayesian inference
(and of statistical inference). There is no consensus on what to do when
the Bayes rule is not applicable.

Machine learning\cite{DeepCritical}
(including numerical analysis in
general\cite{hennig2022probabilistic}) and Quantum
Mechanics\cite{Streater2000Classical}
can be defined as a particular case of Probability updates. Thus, it is
no surprise that there is a broad consensus that Machine learning and
Quantum Mechanics are a generalization of Bayesian inference in some
sense, but little consensus beyond that (such discussion is beyond the
scope of this article, but it appears in
reference\cite{pedro_aligned_2023}of the
same author).

A Bayesian model\cite{seitz} is defined
by a regular conditional probability between standard measure spaces,
also called a Markov kernel or likelihood. If we concatenate Bayesian
models, such that the data (output) of one Bayesian model are the
parameters (input) of another Bayesian model we create a Markov process.

Markov processes cannot produce an arbitrary function of time, because
there is an ordering (related with the concept of entropy) with respect
to which all continuous-time Markov processes are
monotonic\cite{2010entropy}. Thus, Bayesian inference is irreversible.

In this article we propose reversible statistical models and a method of
statistical inference defined by a quantum time-evolution which is
reversible (we call it Unitary inference). The Bayes rule is applied to
the final likelihood which results from the concatenation of several
reversible models followed by the application of the Born rule.

Thus, such reversible inference is fully compatible with Bayesian
inference, that is: given a Bayesian model, a prior probability and data
that allows to produce a posterior probability through Bayesian
inference, there is a reversible model that produces the same posterior
probability as a function of the prior probability (see next section).
Moreover, a quantum time-evolution can be used to define Probability
updates, Machine Learning and Quantum Mechanics.

The crucial advantage of Unitary inference compared to Bayesian
inference, besides the fact that it is more general, is that any prior
wave-function can be transformed into any other prior wave-function
through a change of basis, which also affects the unitary statistical
models. In Bayesian inference, there are priors which are too wide, in
the sense that the relevant sample space is too big to allow for a
numerical integration. At first sight, unitary inference makes this
problem worse because instead of one numerical integration we have to do
many numerical integrations because we are dealing with matrices now.
But it turns out that it also allows for basis changes and thus a prior
wave-function that is too wide in one basis, can become extremely simple
in another basis. Since, the basis change also affects the unitary
statistical models, it is only the relation between the unitary
statistical model and the prior probability distribution that is
relevant. In any case, any approximation (Monte-Carlo, Neural networks,
variational inference, etc.) can be mapped to the formalism of unitary
inference (because it is the most general formalism), where the
approximation can be seen as the choice of a favorable basis and prior
probability distribution with respect to an approximation (through
truncation, for instance) of the unitary statistical model.

Note that probabilities can always be seen as incomplete information,
which in turn can be measured by entropy as defined in classical
information theory. We claim that reversible models and unitary
inference are not in themselves a source of entropy, but they still must
deal with the entropy coming from other sources. This is guaranteed by
the fact that unitary inference is fully compatible with Bayesian
inference. Thus, we do not claim that unitary inference eliminates the
entropy that comes from other sources.

\hypertarget{any-conditional-probability-measure-in-a-standard-measure-space-is-parametrized-by-a-unitary-operator}{%
\section{3. Any conditional probability measure in a standard measure
space is parametrized by a unitary
operator}\label{any-conditional-probability-measure-in-a-standard-measure-space-is-parametrized-by-a-unitary-operator}}

Measure spaces (in particular, algebraic measure theory as defined
below) have many applications in classical mechanics and dynamical
systems\cite{mauroy2020koopman}, when non-linear equations and/or complex sample spaces
become an obstacle.

Any measure space can be defined as a probability space: a probability
space is a measure space where a probability density was chosen, which
is any measurable function normalized such that its measure is 1. In the
historical (that is, Kolmogorov's) probability theory, a probability
space has three parts: a sample space (which is the set of possible
states of a system, also called phase space); a complete Boolean algebra
of events (where each event is a subset of the set of possible states);
and a probability measure which assigns a probability to each event. The
probability is a map from complex random events to abstract random
events, shifting all ambiguity with the notion of randomness to the
abstract random events as described in the following: that the
probability of an event is \(0.32\) means that our event has the same
likelihood of finding a treasure that we know it was hidden in the sand
of a 1 km wide beach, if we only search for it with a metal detector in
a 320~m interval. While this treasure hunt is ambiguous (are there any
clues for the location of the treasure? Etc.) the map from our complex
events to this treasure hunt is unambiguous.

On the other hand, a \emph{standard} measure space is isomorphic (up to
sets with null measure) to the real Lebesgue measure in the unit
interval or to a discrete (finite or countable) measure space or to a
mixture of the two. Thus, topological notions such as dimension do not
apply to \emph{standard} measure spaces. Most probability spaces with
real-world applications are \emph{standard }measure spaces.
Equivalently, a \emph{standard} measure space can be defined such that
the following correspondence holds: every commutative von Neumann
algebra on a separable (complex or real) Hilbert space is isomorphic to
\(L^\infty(X,\mu)\) for some standard measure space \((X,\mu)\) and
conversely, for every standard measure space~\((X,\mu)\) the
corresponding Lebesgue space \(L^\infty(X,\mu)\)~is a von Neumann
algebra. As expected, the representation of an algebra of events in a
(real or complex) Hilbert space uses projection-valued
measures\cite{Pedro2013On}\cite{Moretti2016Quantum}\cite{Li2003Real}\cite{Moretti2013Spectral}. A
projection-valued measure assigns a self-adjoint projection operator to
each event, in such a way that the boolean algebra of events is
represented by the commutative von Neumann algebra: intersection/union
of events is represented by products/sums of projections, respectively.
The state of the ensemble is a linear functional which assigns a
probability to each projection. Thus, there is an algebraic measure
theory\cite{mauroy2020koopman}\cite{whittle2000probability}\cite{aps}\cite{algprob} based on
commutative von Neumann algebras on a separable Hilbert space which is
essentially equivalent to measure theory (for standard measure spaces).

To be sure, the algebraic measure theory is based on \emph{commutative}
algebras, thus it is not a non-commutative generalization of probability
or information theory (see
\protect\hyperlink{quantum-mechanics-versus-a-non-commutative-generalization-of-probability-theory}{section
4}). That is, there is no need for a conceptual or foundational
revolution such as qubits replacing bits when switching from the
historical to the algebraic probability theory
\cite{whittle2000probability}.
Moreover, this is a common procedure in mathematics, as illustrated in
the following quote \cite{integration} (note that a probability measure is related to integration):

\begin{quote}
\emph{The fundamental notions of calculus, namely differentiation and
integration, are often viewed as being the quintessential concepts in
mathematical analysis, as their standard definitions involve the concept
of a limit. However, it is possible to capture most of the essence of
these notions by purely algebraic means (almost completely avoiding the
use of limits, Riemann sums, and similar devices), which turns out to be
useful when trying to generalize these concepts}{[}...{]}

T. Tao (2013)\cite{integration}
\end{quote}

The relation of the algebraic measure theory with probability theory is
the following: any joint probability density \(p(x,y)\) between two
\emph{standard} measure spaces \(X, Y\) (possibly with continuous and
discrete parts) can be written as \(p(x,y)=|\mathcal{U}(y,x,0)|^2\)
where \(\mathcal{U}:L^2(\mathbb{Z})\to L^2(X\times Y)\) is a unitary
operator between Hilbert spaces of complex square-integrable functions.
Note that this result is not surprising and can be seen as a commutative
version of Wigner's theorem\cite{Geh_r_2014}. The proof follows.

We have \(p(y|x)p(x)=p(x,y)=|\mathcal{U}(y,x,0)|^2=p(x|y)p(y)\) up to
sets with null measure, for any probability density \(p(x)\) and some
unitary \(\mathcal{U}\) on the separable Hilbert space. Note that in a
standard measure spaces it is always possible to define regular
conditional
probabilities\cite{durrett2019probability} and to choose \(p(x)=p_0(x)>0\) for all \(x\in X\), except in
sets with null measure. Moreover, the choice of \(p_0\) is independent
of \(p(y|x)\). Thus, we will parametrize \(p(x, y)=p(y|x)p_0(x)\) the
joint probability density for the tensor product \((x, y)\in X\times Y\)
for a particular \(p(x)=p_0(x)>0\) for all \(x\in X\), except in sets
with null measure. Then, we can obtain a parametrization for any other
\(p(x)\) from the expression \(p(y|x)p(x)=\frac{p(x,y)}{p_0(x)}p(x)\).

Since \(p(x, y)\geq 0\) then there is always a normalized wave-function
\(\Psi\in L^2(X\times Y)\) such that \(|\Psi(x,y)|^2=p(x\otimes y)\).
Let \(\{e_j\}\) be an orthonormal basis of the Hilbert space
\(L^2(X\times Y)\). Then, we can build the unitary \(U\) through the
Gram-Schmidt process, such that \(\mathcal{U}(y,x,0)=\Psi(x,y)\).

Note that the Cauchy-Schwarz inequality implies
\(|\int dx \Psi(x,y)\Phi(x) |^2\leq \int dx |\Psi(x,y)|^2\) where the
\(L^2\) norm of \(\Phi(x)\) is 1. Thus,
\(||\Psi\{\Phi\}||=\int dy |\int dx \Psi(x,y)\Phi(x) |^2\leq \int dxdy |\Psi(x,y)|^2=1\),
this implies \(\Psi(x,y)\) is bounded when defined as an operator
\(\Psi:L^2(X)\to L^2(Y)\).

Since \(\Psi\) is bounded, then it admits a Singular Value Expansion
(through the polar decomposition), that is, \(\Psi=V D U^\dagger\),
where \(D\) is a positive semi-definite diagonal operator, \(U\) is
unitary and \(V\) is a partial isometry (with kernel where \(D\) is null
and an isometry where \(D\) is non-null).

Then \(\Psi p_0 \Psi^\dagger=\Psi \Psi^\dagger=V D^2 V^\dagger\), but in
general we have
\(T p T^\dagger=\Psi \frac{p}{p_0}\Psi^\dagger=V D U^\dagger \frac{p}{p_0} U D V^\dagger\).

Since the Hilbert space is separable, there is an orthonormal discrete
basis, and we can then enlarge the discrete part of the sample space
\(X\) to include the orthonormal elements \(e_k\) corresponding to the
elements of the basis \(\varphi_k\) that are missing, setting
\(D_{k k}=0\) for the new indices, replacing a would-be partial isometry
\(V\) by the unitary \(W\) built through the Gram-Schmidt process such
that \(VV^\dagger W=V\). We get \(\Psi=W D U^\dagger\).

The converse also holds. Given a bounded operator \(B\), such that
\(tr(BB^\dagger)=1\), then it defines a joint probability distribution
of initial and final states \(p(x, y)=|B(y,x)|^2\). From the joint
probability, if \(p(x)=\{B^\dagger B\}(x,x)>0\) for all \(x\in X\), then
we can define a regular conditional probability density (since the
measure space is standard).

The linearity of the commutative algebra; avoiding a fixed sample space
\emph{a priori}; and the fact that we can map complex random phenomena
to an abstract random process unambiguously\emph{,} are obvious
advantages for algebraic measure theory when we want to compare
probability theory with Quantum Mechanics, where the linearity of the
canonical transformations is guaranteed by Wigner's theorem (it is
essentially a consequence of the Born's rule applied to a
non-commutative algebra of
operators\cite{Geh_r_2014}\cite{Ruxe4tz1996On}\cite{Molnuxe1r1998algebraic});
the Hilbert space of wave-functions replaces the sample space; and the
canonical transformations are non-deterministic.

The algebraic measure theory is also different from defining the sample
space as a reproducing kernel Hilbert space
\cite{fukumizu2013kernel}\cite{10.5555ux2f3023638.3023648}, since no sample space (whether it is a
Hilbert space or not) is defined \emph{a priori}. Note that defining the
sample space as a Sobolev Hilbert space is common in classical field
theory\cite{leoni2009first}, but defining
a general probability measure in such space is still an open problem.

\hypertarget{quantum-mechanics-versus-a-non-commutative-generalization-of-probability-theory}{%
\section{4. Quantum Mechanics versus a non-commutative generalization of
probability
theory}\label{quantum-mechanics-versus-a-non-commutative-generalization-of-probability-theory}}

The correspondence between geometric spaces and commutative algebras is
important in algebraic geometry\footnote{The correspondence between
  geometric spaces and commutative algebras is consequence of the
  Gelfand representation: there is an isomorphism between a commutative
  {\emph{C}\textsuperscript{*}}-algebra {\emph{A}} and the algebra of
  continuous functions of the spectrum of {\emph{A}}.}. It is usually
argued that the phase space in quantum mechanics corresponds to a
non-commutative algebra, and thus it is a non-commutative geometric
space in some
sense\cite{Connes1995Noncommutative}. It
is a fact that Quantum Mechanics may inspire a non-commutative
generalization of probability theory, since the wave-function could also
assign a probability to non-diagonal projections, these non-diagonal
projections would generate a non-commutative
algebra\cite{Streater2000Classical}.
However, after the wave-function collapse, only a commutative algebra of
operators remains. Thus, the phase space in quantum mechanics is a
sample space of a standard measure space and the standard spectral
theory (where the correspondence between geometric spaces and
commutative algebras plays a main
role~\cite{Steen1973Highlights})
suffices.

Consider for instance the projection \(P_X\) to a region of space \(X\)
and a projection \(U P_p U^\dagger\) to a region of momentum \(p\),
where \(P_X\) and \(P_p\) are diagonal in the same basis. The
projections \(P_X\) and \(U P_p U^\dagger\) are related by a Fourier
transform \(U\) and thus are diagonal in different basis and do not
commute (they are complementary observables). Since we can choose to
measure position or momentum, it seems that Quantum Mechanics is a
non-commutative generalization of probability
theory\cite{Streater2000Classical}.

But due to the wave-function collapse, Quantum Mechanics is not a
non-commutative generalization of probability theory despite the
appearances: the measurement of the momentum is only possible if a
physical transformation of the statistical ensemble also occurs, as we
show in the following.

Suppose that \(E(P_X)\) is the probability that the system is in the
region of space \(X\), for the state of the ensemble \(E\) diagonal
(i.e.~verifying \(E(O)=0\) for operators \(O\) with null diagonal).
Using the notation of~\cite{2020Density}, we have \(E(A)=\mathrm{tr}(\rho A)\) where \(A\) is any operator
and \(\rho\) is an (eventually unbounded) self-adjoint operator with
\(\mathrm{tr}(\rho)=1\) and \(\rho\) is diagonal in the same basis where
the projection operators \(P_X\) are diagonal. If an operator \(O\) has
null diagonal in the same basis where \(P_X\) is diagonal, then
\(tr(\rho O)=0\) for any \(\rho\) diagonal.

If we consider a unitary transformation \(U\) on the ensemble, then
after the wave-function collapse\cite{Brun_2002} we have a new ensemble with state \(E_U\) given by:

\begin{equation}\begin{aligned}
    E_U(A)=&tr(\rho_U A)\end{aligned}\end{equation}

Where \(\rho_U\) is diagonal (due to the wave-function collapse) in the
same basis where the projection operators \(U P_p U^\dagger\) are
diagonal, defined such that:

\begin{equation}\begin{aligned}
    E_U(D)&= \mathrm{tr}(\rho_U D)=\\
    &=\mathrm{tr}(\rho U D U^\dagger)=E(U D U^\dagger)\end{aligned}\end{equation}

Where \(D\) is any diagonal operator and \(O\) is an operator with null
diagonal in the same basis where \(P_X\) and \(P_p\) are diagonal. Thus,
due to the wave-function collapse:

\begin{equation}\begin{aligned}E_U(O)&=0\end{aligned}\end{equation}

Thus, \(E_U(P_p)=E(U P_p U^\dagger)\) is the probability that the system
is in the region of momentum \(p\), for the state of the ensemble
\(E_U\). But the ensembles \(E\) and \(E_U\) are different, there is a
physical transformation relating them.

Without collapse, we would have \(E_U(O)=E(U O U^\dagger)\neq 0\) for
operators \(O\) with null-diagonal, and we could talk about a common
state of the ensemble \(E\) assigning probabilities to a non-commutative
algebra. But the collapse keeps Quantum Mechanics as a Kolmogorov's
probability theory, even when complementary observables are considered.
We could argue that the collapse plays a key role in the consistency of
the theory, as we will see below.

At first sight, our result that the wave-function is merely a
parametrization of any probability measure, resembles Gleason's
theorem~\cite{Gleason1957Measures}\cite{Dvurecenskij2013Gleasonux2019s}. However, there is a key difference: we are dealing with
commuting projections and consequently with the wave-function, while
Gleason's theorem says that any probability measure for all
\emph{non-commuting} projections defined in a Hilbert space (with
dimension \(\geq 3\)) can be parametrized by a density matrix. Note that
a density matrix includes mixed states, and thus it is more general than
a pure state which is represented by a wave-function.

We can check the difference in the 2-dimensional real case. Our result
is that there is always a wave-function \(\Psi\) such that
\(\Psi^2(1)=\cos^2(\theta)\) and \(\Psi^2(2)=\sin^2(\theta)\) for any
\(\theta\).

However, if we consider non-commuting projections and a diagonal
constant density matrix \(\rho=\frac{1}{2}\), then we have:

\begin{equation}\begin{aligned}
\begin{cases}
\mathrm{tr}(\rho \left[\begin{smallmatrix} 1 & 0\\ 0 & 0 \end{smallmatrix}\right])=\frac{1}{2}\\
\mathrm{tr}(\rho \frac{1}{2}\left[\begin{smallmatrix} 1 & 1\\ 1 & 1 \end{smallmatrix}\right])=\frac{1}{2}
\end{cases}\end{aligned}\end{equation}

Our result implies that there is a pure state, such that:

\begin{equation}\begin{aligned}
\mathrm{tr}(\rho \left[\begin{smallmatrix} 1 & 0\\ 0 & 0 \end{smallmatrix}\right])=\frac{1}{2}\end{aligned}\end{equation}

(e.g.
\(\rho=\frac{1}{2}\left[\begin{smallmatrix} 1 & 1\\ 1 & 1 \end{smallmatrix}\right]\))

And there is another possibly different pure state, such that:

\begin{equation}\begin{aligned}
\mathrm{tr}(\rho \frac{1}{2}\left[\begin{smallmatrix} 1 & 1\\ 1 & 1 \end{smallmatrix}\right])=\frac{1}{2}\end{aligned}\end{equation}

(e.g.
\(\rho=\left[\begin{smallmatrix} 1 & 0\\ 0 & 0 \end{smallmatrix}\right]\))

But there is no \(\rho\) which is a pure state, such that:

\begin{equation}\begin{aligned}
\begin{cases}
\mathrm{tr}(\rho \left[\begin{smallmatrix} 1 & 0\\ 0 & 0 \end{smallmatrix}\right])=\frac{1}{2}\\
\mathrm{tr}(\rho \frac{1}{2}\left[\begin{smallmatrix} 1 & 1\\ 1 & 1 \end{smallmatrix}\right])=\frac{1}{2}
\end{cases}\end{aligned}\end{equation}

On the other hand, Gleason's theorem implies that there is a \(\rho\)
which is a mixed state, such that :

\begin{equation}\begin{aligned}
\begin{cases}
\mathrm{tr}(\rho \left[\begin{smallmatrix} 1 & 0\\ 0 & 0 \end{smallmatrix}\right])=\frac{1}{2}\\
\mathrm{tr}(\rho \frac{1}{2}\left[\begin{smallmatrix} 1 & 1\\ 1 & 1 \end{smallmatrix}\right])=\frac{1}{2}
\end{cases}\end{aligned}\end{equation}

Gleason's theorem is relevant if we neglect the wave-function collapse,
since it attaches a unique density matrix to non-commuting operators.
However, the wave-function collapse affects differently the density
matrix when different non-commuting operators are considered, so that
after measurement the density matrix is no longer unique. In contrast,
without the wave-function collapse, the wave-function parametrization of
a probability measure would not be possible.

Another difference is that our result applies to standard probability
theory, while Gleason's theorem applies to a non-commutative
generalization of probability theory.

\hypertarget{free-field-parametrization-in-bayesian-inference-and-statistical-mechanics}{%
\section{5. Free field parametrization in Bayesian inference and
Statistical
Mechanics}\label{free-field-parametrization-in-bayesian-inference-and-statistical-mechanics}}

In Bayesian inference there is always a prior probability distribution,
and there is no prior which is better for all
cases\cite{dutchbook}: we
always have to make assumptions. For instance, if we choose a uniform
prior in a continuous sample space then the maximum likelihood coincides
with the maximum of the posterior (resulting from the prior once the
data is taken into account). However, such maximum has null measure and
thus no particular meaning. If we take a sample based in the posterior,
we expect the sample to be somewhere near the maximum but never exactly
at the maximum. Overfitting means that the inference process produced a
sample which is inconsistent with our prior beliefs and thus could not
be produced by Bayesian inference with an appropriate prior. Thus,
overfitting means we need to choose another prior more consistent with
our prior beliefs.

In Bayesian inference, the likelihood of the output data including
correlations and variances fully determines the statistical model; then
all statistical models can be seen as a particular case of one general
statistical model for particular prior knowledge about the parameters of
the general statistical
model\cite{https:ux2fux2fdoi.orgux2f10.48550ux2farxiv.2112.10510}. Thus, there is a probability distribution (the prior) of
a probability distribution (the likelihood of the output data).

But functions (such as the likelihood of the output data) are in general
infinite-dimensional spaces, so it makes sense to look for measures in
infinite-dimensional spaces. While the Lebesgue measure cannot be
defined in a Euclidean-like infinite-dimensional
space\cite{baker1991lebesgue}\cite{baker2004lebesgue}\cite{baker1991lebesgue2}\cite{baker2004lebesgue2}, it is
well known since many decades that a uniform (Lebesgue-like) measure of
an infinite-dimensional sphere can be defined using the Gaussian measure
and the Fock-space (the Fock-space is a separable Hilbert space used in
the second quantization of free quantum
fields)\cite{Peterson2019GaussianLA}.
Such a space can parametrize (we call it the free field parametrization)
the probability distribution of another probability distribution, which
is exactly what we need: the infinite-dimensional sphere parametrizes
the space of all likelihoods of the output data, while the wave-function
whose domain is the sphere parametrizes a measure on the sphere.

In the free field parametrization, the uniform prior over the sphere
defines a vector of the Hilbert space which when used as the prior for
Bayesian inference with arbitrary data generates an orthogonal basis for
the whole Fock-space. Such basis is related with a point process, with
the number of points with a given feature corresponding to the number of
modifications to the uniform prior (in the part of the sample space
corresponding to such feature). Since Bayesian inference with any other
prior can be seen as a combination of the results of different Bayesian
inferences with the uniform prior for different data (eventually an
infinite amount of data for the cases with null measure), then the
uniform prior in the free field parametrization is in many cases (not in
all cases\cite{dutchbook})
appropriate for Bayesian inference in the absence of any other
information.

Moreover, the frequentist view of probabilities is viable because many
statistical problems can be solved by a Bayesian model where the
parameters defining probabilities are inferred from Binomial processes
with uniform prior (for instance, the Binomial converges to the Normal
distribution under some conditions), which implies that it is often
possible to incorporate prior knowledge efficiently by modifying the
counting of events. This does not exclude of course just using an
arbitrary prior (with methods such as preconditioned Crank--Nicolson
MCMC for high-dimensional problems), but that might be computationally
inefficient in some cases. In both cases (counting or MCMC), the
free-field parametrization is crucial to the implementation of many
constrained problems, this allows the uniform prior to be used in many
cases where we are just interested in a subset of the sample space with
null measure (as we discuss elsewhere).

Thus, ensemble forecasting\cite{wu2021theory}---with many applications and where an ensemble of different
statistical models is built---can be seen as sampling from a Bayesian
posterior corresponding to a particular Bayesian prior which selects
which models constitute the ensemble.

This leads us to classical statistical mechanics: whatever system we
study, we need a probability measure on the phase-space of such system
corresponding to an ensemble which defines a Bayesian prior. When the
system we are studying is itself an ensemble, and thus it is defined by
another probability distribution, then we can use the free field
parametrization.

Quantum statistical mechanics is not different, since the Hilbert space
on which the density matrix lives is merely a parametrization of a
probability, due to the wave-function collapse. When the density matrix
is not pure, then the probability defining the ensemble is a joint
probability distribution of the initial and final states of the systems.
We can always define the density matrix through a diagonal operator
rotated by a unitary operator, with the diagonal operator defining the
marginal probability of the initial state and the unitary operator
defining the conditioned probability of the final state conditioned by
the initial state.

We can also consider a statistical model which acts in the prior
wave-function in a non-linear way, by redefining the Hilbert-space of
the prior as a tensor product of two Fock-spaces and the corresponding
statistical model as unitary (and thus linear), this is in principle
possible for most non-linear actions we can think of, because a
conditional probability measure includes all such actions. Note that
every formalism has limits, so there will be always some exotic
statistical model which cannot be redefined or at least not in a useful
way.

This allows us to treat (classical or quantum) statistical processes as
classical dynamical systems (where the system is itself an ensemble).

\hypertarget{free-field-parametrization-for-finite-sample-spaces-and-the-spin-statistics-theorem}{%
\section{6. Free field parametrization for finite sample spaces and the
spin-statistics
theorem}\label{free-field-parametrization-for-finite-sample-spaces-and-the-spin-statistics-theorem}}

The free field parametrization defined in the previous section depends
on the fact that the sphere is infinite dimensional corresponding to a
sample space with infinite degrees of freedom. But often we are
interested in a tensor product of sample spaces, some of which have
finite degrees of freedom.

Of course, we can always treat finite degrees of freedom as a subset of
infinite degrees of freedom. But there might be an alternative as well.

We consider the sample space \(Z_2\) with \(2\) degrees of freedom and
fermionic creation and destruction operators for the pair of degrees of
freedom. Then, the corresponding Fermionic Fock-space parametrizes the
probability space.

If we consider now the sample space \(Z_2\times Z_2\), then naively
there is a possibility of non-null products of creation operators
corresponding to each pair of degrees of freedom. This is not a free
field parametrization.

We need the space
\(\Gamma^s(L^2(\mathbb{Z}_2))\otimes \Gamma^a(L^2(\mathbb{Z}_2)) \)~\cite{Petz:1990gb}\cite{Fock}. The bosonic free
field corresponds to operators with bosonic number, while the fermionic
free field corresponds to operators with fermionic number, without the
need for products of free fields. Instead of \(Z_2\times Z_2\) we could
easily repeat the exercise for \(Z_2^n\times \mathbb{R}^m\).

The advantage is that there are certain symmetry transformations (such
as the rotation for different spins in the presence of a flavor
symmetry\cite{duck1998toward}) which can only be represented by bosonic free fields while
others can only be represented by fermionic free fields. This is what
leads to the spin-statistics
theorem\cite{duck1998toward}
or to the fact that ghost fields verify an unusual spin-statistics
correspondence. So, we better have an alternative to the bosonic free
field parametrization discussed in the previous section.

\hypertarget{symmetries-and-unitary-representations}{%
\section{7. Symmetries and unitary
representations}\label{symmetries-and-unitary-representations}}

Any conditional probability measure (in a standard measure space) which
is function of a one-dimensional parameter (which we call time, without
loss of generality) can be parametrized by a quantum process: a
time-ordered product integral of unitary operators, defined by a
time-dependent Hamiltonian operator, at the cost of choosing a larger
sample space such that the corresponding isometry can be unitary. Such
parametrization using a time-dependent Hamiltonian is particularly
useful in systems where the words ``time-dependent'' have a clear
meaning. For instance, a molecule in an ideal gas is constrained both
theoretically (through its position operator and the effect that the
Poincare group has on it) and experimentally to have conditional
probability measures defined by a time-independent Hamiltonian operator;
also, in a deterministic time-evolution, a time-independent classical
Hamiltonian corresponds to a time-independent Hamiltonian operator.
Nevertheless, such parametrization is always possible in a standard
measure space, including for Markov processes; therefore quantum
processes are not intrinsically exotic or weird, they are simply more
general than Markov processes.

Moreover, as it was discussed in the previous two sections, a non-linear
transformation of the wave-function can be redefined as a unitary
operator (and thus linear) on a Fock-space. This is in principle
possible for most non-linear transformations we can think of, because
the Hilbert space is infinite-dimensional.

Note that Markov processes cannot produce an arbitrary function of time,
because there is an ordering (related with the concept of entropy) with
respect to which all continuous-time Markov processes are
monotonic\cite{2010entropy}. Moreover, a Markov process is a semigroup which is much harder
to translate into an infinitesimal structure than the unitary operators
which form a group, as discussed in
reference\cite{hilgert2006lie}:

\begin{quote}
``\emph{The basic feature of Lie theory is that of using the group
structure to translate global geometric and analytic problems into local
and infinitesimal ones. These questions are solved by Lie algebra
techniques which are essentially linear algebra and then translated back
into an answer to the original problem. Surprisingly enough it is
possible to follow this strategy to a large extend also for semigroups,
but things become more intricate. Because of the missing inverses one
has not only to deal with linear algebra but also with convex geometry
at the infinitesimal level.}''
\end{quote}

A quantum process can be classified by all unitary transformations of
the wave-function that may occur at each time (called canonical
transformations). When a subset of canonical transformations forms a
group, such subset is always a linear unitary representation of a group
which is then called symmetry group and the canonical transformations
are then called symmetry transformations.

In conclusion, conditional probability measures defined by a linear
unitary representation of a symmetry group are certainly natural and
common.

\hypertarget{conservative-transformations}{%
\section{8. Conservative
transformations}\label{conservative-transformations}}

A conservative transformation \(U\) is a canonical transformation that
preserves the measure \(\mu\) of the commutative von Neumann algebra
\(L^\infty(X,\mu)\). Note that the measure \(\mu\) is an element of the
dual space of \(L^\infty(X,\mu)\) and thus an element of the spectrum of
\(U\), but not necessarily of the space \(L^1(X,\mu)\) pre-dual of
\(L^\infty(X,\mu)\).

If there is a locally compact abelian group \(G\) with a transitive free
action on the space \(X\) and a conservative transformation \(U\)
conserves the equation \(G=1\). Then there is a unitary time-evolution
\(U(t)=e^{iH t}\) with a self-adjoint operator \(H\) such that
\([[H,P_A],P_B]=0\) where \(A,B\subset X\) are any events and
\(P_A\in  L^\infty(X,\mu)\) is a projection-valued measure, which is
conservative. We just need to write
\(H=\sum\limits_{j=1}^{n}p_j[H,x_j]+[H,x_j]p_j\), then:

\begin{equation}U(t)g U^\dagger(t)|_{G=1}=1\end{equation}

For all \(g\in G\). Note that \(|_{G=1}\) is a right ideal of the
algebra of bounded operators on the Hilbert space and
\([x_j,p_k]=i\delta_{jk}\), \([x_j,P_A]=0\) and \(p_j\) are the
generators of the abelian group \(G\). The operator \(H\) is uniquely
defined by \([H,x_j]\) up to a constant.

In particular, a deterministic automorphism (defined in the next
section) can be made conservative without modifying its effect on
\(L^\infty(X,\mu)\), if there is a locally compact abelian group \(G\)
with a transitive free action on \(X\). In that case, the deterministic
conservative automorphism can be seen as just a change of variables.
Note that any unitary automorphism can be defined as a time-evolution,
for some Hamiltonian and time. However, a conservative time-evolution is
not necessarily deterministic.

A particular case of a deterministic conservative automorphism is a
measure-preserving transformation \(U\), defined such that \(U\) is a
change of variables with unit Jacobian.

\hypertarget{deterministic-transformations}{%
\section{9. Deterministic
transformations}\label{deterministic-transformations}}

Crucially, the symmetry transformations include all the deterministic
transformations, which will be defined in the following. Thus, the
symmetry transformations are a generalization of the deterministic
transformations.

A deterministic transformation \(P_A\to U^\dagger P_A U\) where \(P_A\)
is a projection operator and \(A\subset X\) is any event, is such that
\(L^\infty\to U L^\infty(X,\mu)U^\dagger=L^\infty(Y,\nu)\), that is all
(necessarily diagonal) elements of the abelian algebra are transformed
in diagonal elements of another abelian algebra. Then, the corresponding
conditional probability density is diagonal, and it is the
Radon--Nikodym derivative~of the measure \(\mu\) with respect to
\(\nu\)\cite{takesaki2012theory}.

In the particular case of a one-parameter continuous group, the
deterministic transformations must be automorphisms. Thus, a
deterministic automorphism \(P_A\to U^\dagger P_A U\) is such that
\(U^\dagger P_A U\) commutes with \(L^\infty=L^\infty(X,\mu)\), where
\(P_A\) is a projection operator and \(A\subset X\) is any event. Since
\(L^\infty(X,\mu)\) is maximal abelian, and it is a commutative unital
C*-algebra\cite{Blecher_2022} then \(U^\dagger P_A U\in L^\infty\) can be written
as a function of the spectrum of \(L^\infty\) (Gelfand representation).
Such function is invertible because any element of the spectrum of
\(L^\infty\) can be written as a function of the spectrum of
\(U^\dagger L^\infty U\) and the other way around. Thus, the points of
the spectrum of \(L^\infty\) are mapped one-to one to other points of
the spectrum of \(L^\infty\), irrespective of the probability
distribution. Since we have effectively a deterministic case, then we
call deterministic transformation to such transformation.

We conclude that an automorphism \(U\) is deterministic if and only if
\(P_A\) and \(U P_B U^\dagger\) commute for all events \(A, B\). Thus,
the complementarity of two observables (e.g.~position and momentum) is
due to the random nature of the symmetry transformation relating the two
observables. This clarifies that probability theory has no trouble in
dealing with non-commuting observables, as long as the collapse of the
wave-function occurs. Note that Quantum Mechanics is not a
generalization of probability theory, but it is definitely a
generalization of classical statistical mechanics since it involves
non-deterministic symmetry transformations. For instance, the time
evolution may be non-deterministic unlike in classical statistical
mechanics.

\hypertarget{ensemble-forecasting-allows-the-approximation-of-a-non-linear-infinite-dimensional-model-by-a-direct-sum-of-linear-models-with-few-variables}{%
\section{10. Ensemble forecasting allows the approximation of a
non-linear infinite-dimensional model by a direct sum of linear models
with few
variables}\label{ensemble-forecasting-allows-the-approximation-of-a-non-linear-infinite-dimensional-model-by-a-direct-sum-of-linear-models-with-few-variables}}

\begin{quote}
\emph{The miracle of the appropriateness of the language of mathematics
for the formulation of the laws of physics is a wonderful gift which we
neither understand nor deserve. ---} Eugene Wigner (1960)
\cite{https:ux2fux2fdoi.orgux2f10.1002ux2fcpa.3160130102}
\end{quote}

There is no uniform countable measure. Thus, the rationals are not
enough for Probability Theory. A standard probability space (which has
countable and continuous measures) seems irreducible.

However, the rules to update the probability measure are not at all
obvious, because we can always consider a probability space of
probability spaces (ensemble forecasting). Assuming a quantum
time-evolution (that is, linear) for the ensemble forecasting leads to
no relevant restriction on the time-evolution of the probability spaces
(it may be non-linear, although every formalism has a domain of
application, so we still need to exclude a pathological time-evolution).

The Bayes rule applies only when we do a measurement, but this is a
tautology. Since a measurement is not defined, then a measurement is
applying the Bayes rule to update a probability: destroy the uncertainty
(and thus the probability distribution) by converting probabilities into
events, unpredictably and reproducing the probability distribution in
the limit of infinite measurements. The limit of infinite measurements
creates a joint probability distribution, with the results of the
infinite measurements and the probability update after each measurement
took place. It is not different from an unbiased sampling process taken
over infinite time. Note that slightly biased sampling processes are
common, but these can be converted to unbiased by enlarging the
probability space (using ensemble forecasting, for instance).

Using the free field parametrization of ensemble
forecasting\cite{wu2021theory} (see Sections
5, 6), we can diagonalize the quantum time-evolution, decomposing a
non-linear infinite-dimensional model in a direct integral of linear
models with only one boolean variable. This only solves part of the
problem, since a direct integral can still be uncomputable. We have to
show how to decompose an eventually continuous spectrum of the quantum
time-evolution into a direct sum of sufficiently small intervals of
energy, each of these intervals described by few variables. This is not
obvious, since any continuous interval, no matter how small, is still
uncountable. We will prove this in another article. But it is not
unreasonable to assume it is possible to do it (for instance, adapting
the Krylov subspace methods for computations of matrix
exponentials\cite{wang2017error}),
as we do from now on.

Since the wave-function defines an ensemble of deterministic
transformations (that is, functions), we can easily study and calculate
each possible function when we are dealing with few variables. This
justifies why deterministic logic and deterministic mathematics find
application in a world full of uncomputable functions.

This gives a mathematical definition on the renormalization process,
which has been empirically observed to be possible to do in a wide range
of problems (related or unrelated to
Physics)\cite{Machta_2013}\cite{B_ny_2015}. That
is, for reasonable initial conditions (because we do not have access to
an infinite range of energies in the real-world, using man-made
experimental apparatus) we can always approximately predict the
time-evolution of a system using a model with few variables, using
different unrelated models for different energy ranges.

The renormalization process is universal, and it applies to any
statistical model (which defines a unitary transformation, playing the
role of a quantum time-evolution, the ``time'' is abstract here). Thus,
it allows efficient machine learning using Bayesian priors in the few
most relevant variables, helping align models and incorporate prior
knowledge.

\hypertarget{conditions-for-the-classical-limit-of-quantum-mechanics}{%
\section{11. Conditions for the classical limit of Quantum
Mechanics}\label{conditions-for-the-classical-limit-of-quantum-mechanics}}

In the previous section, the few variables are related with the
energy-momentum space, not with the coordinate space. Here we discuss
the evidence about the necessary conditions to discard probabilities in
the coordinate space, recovering Classical Mechanics. Note that these
conditions are not rigorous, that is, often it works but not always,
thus Classical Mechanics is never enough at any energy scale, Quantum
Mechanics is always needed.

As far as we know the world is compatible with a continuum space-time,
with rational functions of rational variable obtained through step
functions and the partition of a continuum space-time. For instance, we
can ``see'' that we have ``five'' fingers in our hand, but exactly why
we can do this in a continuum space-time is not obvious. The mainstream
mathematics depends on the fact that step, polynomials and smooth
functions are relevant for applications. But it is important to look to
the existing evidence on why these functions are relevant for
applications.

Step, polynomial or smooth functions are dense in the \(L^2\) measure.
Any bounded function in a compact domain, can be approximated in the
\(L^2\) measure (that is, the average over a compact domain, which is
common to do in classical physics when we partition a large domain in
numerical approximations) by these calculable functions. That does not
make the original function calculable, since often we do not want to
average over a compact domain, specially when studying microscopic
phenomena such as the double slit experiment (where it makes a
macroscopic difference to open one or two microscopic slits for a
particle to go through). This is consistent with the fact that
calculable functions are useful in macroscopic phenomena, but not so
much in microscopic phenomena, where only the probabilities admit
calculable approximations, because they still can be approximated in the
\(L^2\) measure.

Note that Chaos and singularities in Ordinary differential equations are
phenomena of computable functions (computable at least for a small
enough but finite time, thus they still admit the partition of a large
domain in numerical approximations), despite that these are clues for
the fact that many functions are uncomputable. Uncomputable functions
are uncomputable for any finite time, unlike Chaos or singularities.

\hypertarget{refs}{}

\fontsize{8}{7}\selectfont

\newpage{\pagestyle{empty}\cleardoublepage}

\end{document}